\documentclass[11pt,fleqn]{iopart}
\usepackage[utf8]{inputenc}
\usepackage[T1]{fontenc}
\usepackage{fouriernc}
\expandafter\let\csname equation*\endcsname=\relax
\expandafter\let\csname endequation*\endcsname=\relax
\usepackage{amsmath,amssymb,amsfonts,amsthm}
\usepackage{hyperref,xcolor}
\usepackage{etoolbox}
\usepackage{units}

\numberwithin{equation}{section}

\def\sl{{\mathfrak{sl}}}
\def\os{{\mathfrak{o}}}
\def\su{{\mathfrak{su}}}
\def\us{{\mathfrak{u}}}
\def\sp{{\mathfrak{sp}}}

\newcommand{\olin}[1]{\overline{#1}}
\newcommand{\wh}[1]{\widehat{#1}}

\begin{document}

\title[The Higgs and Hahn algebras from a Howe duality perspective]{The Higgs and Hahn algebras from a Howe duality perspective}

\author{Luc Frappat}
\ead{luc.frappat@lapth.cnrs.fr}
\address{Laboratoire d'Annecy-le-Vieux de Physique Th\'eorique LAPTh, Univ. Grenoble Alpes, Univ. Savoie Mont Blanc, CNRS, F-74000 Annecy, France}

\author{Julien Gaboriaud}
\ead{gaboriaud@CRM.UMontreal.CA}
\address{Centre de Recherches Math\'ematiques, Universit\'e de Montr\'eal, Montr\'eal (QC), Canada}

\author{Luc Vinet}
\ead{vinet@CRM.UMontreal.CA}
\address{Centre de Recherches Math\'ematiques, Universit\'e de Montr\'eal, Montr\'eal (QC), Canada}

\author{St\'ephane Vinet}
\ead{stephanevinet@uchicago.edu}
\address{The College, The University of Chicago, 5801 S. Ellis Ave, Chicago, IL 60637, USA}

\author{Alexei Zhedanov}
\ead{zhedanov@yahoo.com}
\address{Department of Mathematics, Renmin University of China, Beijing 100872, China}

\vspace{10pt}

\begin{abstract}
The Hahn algebra encodes the bispectral properties of the eponymous orthogonal polynomials. In the discrete case, it is isomorphic to the polynomial algebra identified by Higgs as the symmetry algebra of the harmonic oscillator on the $2$-sphere. These two algebras are recognized as the commutant of a $\mathfrak{o}(2)\oplus\mathfrak{o}(2)$ subalgebra of $\mathfrak{o}(4)$ in the oscillator representation of the universal algebra $\mathcal{U}(\mathfrak{u}(4))$. This connection is further related to the embedding of the (discrete) Hahn algebra in $\mathcal{U}(\mathfrak{su}(1,1))\otimes\mathcal{U}(\mathfrak{su}(1,1))$ in light of the dual action of the pair $\big(\mathfrak{o}(4),\mathfrak{su}(1,1)\big)$ on the state vectors of four harmonic oscillators. The two-dimensional singular oscillator is naturally seen by dimensional reduction to have the Higgs algebra as its symmetry algebra.
\end{abstract}

\section{Introduction}
This paper is concerned with the Higgs algebra \cite{Higgs1979,Letourneau1995,Genest2014} and the discrete version of the Hahn algebra \cite{Granovskii1991,Granovskii1992} which actually designate two different but isomorphic presentations of the same algebra \cite{Genest2014,Zhedanov1992}. We aim to establish that this algebra arises as the commutant of a $\os(2)\oplus\os(2)$ subalgebra of $\os(4)$ in the oscillator representation of the universal algebra $\mathcal{U}(\us(4))$. We will moreover point out that this relation that the Higgs and Hahn algebras have with $\os(4)$ is in duality, in the sense of Howe \cite{Howe1987,Howe1989,Rowe2012}, with the one they are known to have with $\su(1,1)\otimes\su(1,1)$ \cite{Granovskii1988,Zhedanov1993}. Let us start with some background.

The Hahn algebra has three generators $\wh{K}_{1}$, $\wh{K}_{2}$ and $\wh{K}_{3}$ subjected to the relations
\begin{align}\label{eq:hahnd}
\begin{aligned}{}
 [\wh{K}_{1},\wh{K}_{2}]&=\wh{K}_{3}\\
 [\wh{K}_{2},\wh{K}_{3}]&=a\{\wh{K}_{1},\wh{K}_{2}\}+b\wh{K}_{2}+c_1\wh{K}_{1}+d_1\\
 [\wh{K}_{3},\wh{K}_{1}]&=a{{\wh{K}_{1}}}{}^{2}+b\wh{K}_{1}+c_2\wh{K}_{2}+d_2
\end{aligned}
\end{align}
where $\{A,B\}=AB+BA$ and $a$, $b$, $c_1$, $c_2$, $d_1$, $d_2$ are structure constants. We assume $a\neq0$ (otherwise \eqref{eq:hahnd} would be equivalent to the Lie algebra $\mathfrak{sl}(2)$). This algebra describes the eigenvalue problems of both the discrete and continuous Hahn polynomials \cite{Koekoek2010}. We shall henceforth consider the discrete case where $c_2<0$ and which is realized by the bispectral operators of the Hahn polynomials (see \cite{Vinet2018} for instance). In this case, upon performing the affine transformation
\begin{align}
\begin{aligned}{}
 \wh{K}_{1}&=\frac{1}{2}\sqrt{-c_2}\,K_1-\frac{b}{2a}\\
 \wh{K}_{2}&=-\frac{1}{2}aK_2-\frac{c_1}{2a}
\end{aligned}
\end{align}
one can cast the commutation relations in the form
\begin{align}\label{eq_Hahn2}
\begin{aligned}{}
 [K_1,K_2]&=K_3\\
 [K_2,K_3]&=-2\{K_1,K_2\}+\delta_1\\
 [K_3,K_1]&=-2{K_1}^{2}-4K_2+\delta_2
\end{aligned}
\end{align}
with $\delta_1$, $\delta_2$ constants (or central elements).

The Hahn algebra admits an embedding in $\mathcal{U}(\mathfrak{su}(1,1))\otimes\mathcal{U}(\mathfrak{su}(1,1))$ that we shall describe in details later as it is germane to our analysis. This observation underscores its connection to the Clebsch-Gordan problem for $\su(1,1)$ (and $\su(2)$).

The Higgs algebra can be viewed as a polynomial deformation of $\su(2)$. It has three generators $D$, $A_+$, $A_-$ satisfying the following commutation relations:
\begin{align}\label{eq_Higgs}
\begin{aligned}{}
 [D,A_\pm]&=\pm4A_\pm\\ 
 [A_+,A_-]&=-D^{3}+\alpha_1D+\alpha_2
\end{aligned}
\end{align}
with $\alpha_1$, $\alpha_2$ central elements. That the Higgs algebra is isomorphic to the discrete Hahn algebra is readily seen by taking
\begin{align}\label{eq_Higgs2Hahn}
\begin{aligned}{}
 K_1&=\frac{1}{2}D\\
 K_2&=-\frac{1}{4}\left(A_++A_-+\frac{1}{2}D^{2}\right)+\frac{\alpha_1}{8}\\
 K_3&=[K_1,K_2]=-\frac{1}{2}(A_+-A_-) 
\end{aligned}
\end{align}
and observing that the commutation relations \eqref{eq_Hahn2} then follow from \eqref{eq_Higgs} with
\begin{align}
 \delta_1=-\frac{\alpha_2}{4}\qquad\delta_2=\frac{\alpha_1}{2}.
\end{align}
Historically, the algebra defined in \eqref{eq_Higgs} was found by Higgs, hence the name, as the one realized by the conserved quantities of the Coulomb problem and harmonic oscillator on the two-sphere. It can be viewed as a deformed $\su(2)$ algebra \cite{Bonatsos1995} or a truncation of the quantum algebra $\mathcal{U}_q(\sl(2))$ \cite{Zhedanov1992}. This algebra has been identified as the symmetry algebra of the Hartmann \cite{Granovskii1991} and of certain ring-shaped potentials \cite{Granovskii1992} as well as the singular oscillator in two dimensions \cite{Letourneau1995,Genest2014}. The Higgs algebra has moreover emerged in the Heisenberg quantization of identical particles \cite{Leinaas1993}. Furthermore, it has been seen to coincide with the finite quantum W-algebra $W(\sp(4),2\sl(2))$ \cite{Bowcock1994,Barbarin1995}. (For a review of finite W-algebras and their applications, see \cite{DeBoer1996}.)

Similarly to the Hahn case, the Racah algebra \cite{Genest2014,Granovskii1988,Genest2014a} is realized by the bispectral operators of the corresponding polynomials. It admits an embedding in $\mathcal{U}(\su(1,1))^{\otimes3}$ with the intermediate Casimir elements representing the generators. The Hahn algebra can be obtained through a contraction of the standard presentation of the Racah algebra in a way that parallels the limit that takes the Racah polynomials into those of Hahn \cite{Koekoek2010}. A generalization of the Racah algebra to higher ranks is found in \cite{DeBie2017a}.

Recently the Racah algebra has been interpreted in a Howe duality framework and shown to be a commutant \cite{Gaboriaud2018} in the enveloping algebra of $\os(6)$, the Lie algebra of the rotation group in six dimensions. An extension of this result to the generalized Racah algebra is given in \cite{Gaboriaud2018a}. An analogous treatment of the Bannai-Ito algebra \cite{Tsujimoto2012,DeBie2015,DeBie2016a}, which is in a sense a supersymmetric version of the Racah algebra, was also achieved in \cite{Gaboriaud2018b}. These advances raised the question of how to describe the Higgs algebra from a Howe duality perspective. The answer to this question will be provided here with the significant merit of expanding and interconnecting the various descriptions of the Higgs and Hahn algebras.

The remainder of the paper is organized as follows. As preparation background, familiar results on the metaplectic representation of $\su(1,1)$ and the embedding of $\us(4)$ in the Heisenberg-Weyl algebra will be reviewed in Section \ref{sec_oscillators}. The Higgs algebra will be obtained as the commutant of $\os(2)\oplus\os(2)$ in $\mathcal{U}(\us(4))$ in Section \ref{sec_Higgs}. The embedding of the Hahn algebra into $\mathcal{U}(\mathfrak{su}(1,1))\otimes\mathcal{U}(\mathfrak{su}(1,1))$ will be described in Section \ref{sec_embedding}. The two pictures of the Higgs/Hahn algebra presented in Sections \ref{sec_Higgs} and \ref{sec_embedding} will be connected via the Howe dual pair $\big(\os(4),\su(1,1)\big)$ that acts on the state vectors of the four-dimensional oscillator. Dimensional reduction will be used in Section \ref{sec_reduction} to recover the fact that the symmetries of the singular oscillator in two dimensions generate the Higgs algebra. The paper will end with a summary of the findings and an outlook.

\section{$\su(1,1)$, $\us(4)$ and oscillators}\label{sec_oscillators}
We shall be dealing with the Heisenberg-Weyl algebra $W(n)$ generated by $n$ pairs of oscillator operators $a_i$, $a_i^{\dagger}$, $i=1,\dots,n$, that satisfy
\begin{align}
 [a_i,a_j^{\dagger}]=\delta_{ij},\qquad i,j=1,\dots,n.
\end{align}
The number operators $N_i=a_i^{\dagger}a_i$ are such that
\begin{align}
 [N_i,a_j]=-a_i\delta_{ij},\qquad [N_i,a_j^{\dagger}]=a_i^{\dagger}\delta_{ij}.
\end{align}
In the position coordinates $x_i$, $i=1,\dots,n$ these operators read
\begin{align}
 a_i=\frac{1}{\sqrt{2}}\left(\frac{\partial}{\partial x_i}+x_i\right),\quad a_i^{\dagger}=\frac{1}{\sqrt{2}}\left(-\frac{\partial}{\partial x_i}+x_i\right),\quad N_i=-\frac{1}{2}\frac{\partial^{2}}{\partial {x_i}^{2}}+\frac{1}{2}{x_i}^{2}-\frac{1}{2}.
\end{align}
The Lie algebra $\su(1,1)$ has generators $J_0$, $J_+$, $J_-$ obeying the commutation relations
\begin{align}
 [J_0,J_\pm]=\pm J_\pm,\qquad [J_+,J_-]=-2J_0.
\end{align}
Its Casimir element is given by
\begin{align}
 C={J_0}^{2}-J_+J_--J_0.
\end{align}
Owing to the fact that $\su(1,1)$ has a trivial coproduct, $J_0^{(12)}=J_0^{(1)}+J_0^{(2)}$, $J_\pm^{(12)}=J_\pm^{(1)}+J_\pm^{(2)}$ with $J_{\bullet}^{(1)}=J_{\bullet}\otimes1$ and $J_{\bullet}^{(2)}=1\otimes J_{\bullet}$, defines an embedding of $\su(1,1)$ into $\su(1,1)\otimes\su(1,1)$. This fact and the notation extend to $\su(1,1)^{\otimes n}$.

The metaplectic representation of $\su(1,1)$ is defined by the following map in $W(1)$:
\begin{align}\label{eq_metaplecticrep}
 \mathcal{J}_0^{(i)}=\frac{1}{2}\left(a_i^{\dagger}a_i+\frac{1}{2}\right),\qquad \mathcal{J}_+^{(i)}=\frac{1}{2}a_i^{\dagger}{}^{2},\qquad \mathcal{J}_-^{(i)}=\frac{1}{2}{a_i}^{2}.
\end{align}
It consists in the direct sum of two irreducible $\su(1,1)$ representations on the spaces spanned respectively by the eigenstates of $N_i=a_i^{\dagger}a_i$ with either even or odd eigenvalues.
The Casimir element $C$ has value $\nicefrac{-3\,\,}{\,\,16}$ in that representation. In the following we shall consider $\mathcal{J}_\bullet^{(1234)}=\mathcal{J}_\bullet^{(12)}+\mathcal{J}_\bullet^{(34)}$ which provides an embedding of $\su(1,1)$ into $W(4)$ as per the remarks above.

The Lie algebra $\us(4)$ with generators $E_{ij}$, $i,j=1,\dots,4$ admits the following realization à la Schwinger in $W(4)$:
\begin{align}
 E_{ij}=a_i^{\dagger}a_j,\qquad i,j=1,\dots,4.
\end{align}
The Hamiltonian of the isotropic harmonic oscillator in four dimensions:
\begin{align}
 H=N_1+N_2+N_3+N_4+2
\end{align}
is central, $[H,a_i^{\dagger}a_j]=0$, and should be excluded from the $16$ independant $a_i^{\dagger}a_j$ to deal with $\us(4)$ per se. In this oscillator representation, the $\os(4)$ subalgebra of $\us(4)$ is spanned by the infinitesimal rotation generators
\begin{align}
 L_{jk}=\frac{i}{2}(a_ja_k^{\dagger}-a_j^{\dagger}a_k)=-\frac{i}{2}\left(x_j\frac{\partial}{\partial x_k}-x_k\frac{\partial}{\partial x_j}\right)
\end{align}
with commutation relations
\begin{align}
 [L_{jk},L_{\ell m}]=\frac{i}{2}(L_{j\ell}\delta_{km}-L_{k\ell}\delta_{jm}+L_{km}\delta_{j\ell}-L_{jm}\delta_{k\ell}),\qquad j,k,\ell,m=1,\dots,4.
\end{align}

\section{The Higgs algebra as a commutant in $\mathcal{U}(\us(4))$}\label{sec_Higgs}
We are now ready to obtain our first main result, namely that the Higgs algebra can be defined as a commutant. Pick the $\os(2)\oplus\os(2)$ subalgebra of $\os(4)$ generated by $L_{12}$ and $L_{34}$; clearly $[L_{12},L_{34}]=0$. We want to concentrate on the commutant of this subalgebra in $\mathcal{U}(\us(4))$. We are thus looking for polynomials in the generators $a_i^{\dagger}a_j$, $i,j=1,\dots,4$ that are invariant under rotations in both the $(1-2)$- and $(3-4)$-planes. It is not difficult to convince oneself that an integrity basis for that set is provided by the three operators
\begin{align}\label{eq_integrity}
\begin{aligned}{}
 A_+&=({a_1^{\dagger}}{}^{2}+{a_2^{\dagger}}{}^{2})({a_3}^{2}+{a_4}^{2})\\
 A_-&=({a_1}^{2}+{a_2}^{2})({a_3^{\dagger}}{}^{2}+{a_4^{\dagger}}{}^{2})\\
 D&=(N_1+N_2)-(N_3+N_4).
\end{aligned}
\end{align}
$A_\pm$ and $D$ are manifestly invariant under the rotations generated by $L_{12}$ and $L_{34}$ and they clearly commute with $H$ (thus belonging to $\mathcal{U}(\us(4))$). All other elements of the commutant are built from those.

Let us now determine the commutation relations of these generators. It is immediate to see that
\begin{align}
 [D,A_\pm]=\pm 4A_\pm.
\end{align}
There remains to evaluate $[A_+,A_-]$. Observe first that one has the following identities:
\begin{align}\label{eq_aid2ai2}
 {a_i^{\dagger}}{}^{2}{a_i}^{2}={N_i}^{2}-N_i
\end{align}
as well as
\begin{align}\label{eq_aid2aj2}
 {a_i}^{2}{a_j^{\dagger}}{}^{2}+{a_i^{\dagger}}{}^{2}{a_j}^{2}=2N_iN_j+N_i+N_j-4{L_{ij}}^{2},\qquad i,j=1,\dots,4.
\end{align}
A straightforward computation yields
\begin{align}
\begin{aligned}{}
 [A_+,A_-]&=4\left(a_1^{\dagger}{}^{2}{a_1}^{2}+a_1^{\dagger}{}^{2}{a_2}^{2}+a_2^{\dagger}{}^{2}{a_1}^{2}+a_2^{\dagger}{}^{2}{a_2}^{2}\right)\left(N_3+N_4+1\right)\\
          &-4\left(N_1+N_2+1\right)\left(a_3^{\dagger}{}^{2}{a_3}^{2}+a_3^{\dagger}{}^{2}{a_4}^{2}+a_4^{\dagger}{}^{2}{a_3}^{2}+a_4^{\dagger}{}^{2}{a_4}^{2}\right)
\end{aligned}
\end{align}
which with the help of \eqref{eq_aid2ai2} and \eqref{eq_aid2aj2} is readily converted to
\begin{align}
\begin{aligned}{}
 [A_+,A_-]&=4\left[(N_1+N_2)^{2}-4{L_{12}}^{2}\right]\left(N_3+N_4+1\right)\\
          &-4\left(N_1+N_2+1\right)\left[(N_3+N_4)^{2}-4{L_{34}}^{2}\right].
\end{aligned}
\end{align}
Since 
\begin{align}
 N_1+N_2=\frac{1}{2}(H+D-2),\qquad N_3+N_4=\frac{1}{2}(H-D-2),
\end{align}
upon substituting and after some algebra, one obtains
\begin{align}
 [A_+,A_-]&=-D^{3}+\left[H^{2}+8\left(L_{12}{}^{2}+L_{34}{}^{2}\right)-4\right]D-8\left(L_{12}{}^{2}-L_{34}{}^{2}\right)H.
\end{align}
Since $H$, $L_{12}$, $L_{34}$ commute with all the generators, we thus conclude comparing with \eqref{eq_Higgs} that indeed the Higgs algebra is the commutant in $\mathcal{U}(\us(4))$ of $\os(2)\oplus\os(2)$ with the structure ``constants'' given by
\begin{align}\label{eq_alpha12}
\begin{aligned}{}
 \alpha_1&=H^{2}+8\left(L_{12}{}^{2}+L_{34}{}^{2}\right)-4\\
 \alpha_2&=-8\left(L_{12}{}^{2}-L_{34}{}^{2}\right)H.
\end{aligned}
\end{align}
This provides a most simple characterization of the Higgs algebra.

We can translate these results in terms of the Hahn presentation. Substituting \eqref{eq_integrity} in \eqref{eq_Higgs2Hahn}, using formula \eqref{eq_aid2aj2} and keeping in mind the expression for $\alpha_1$ given in \eqref{eq_alpha12}, one arrives at the following nice expressions
\begin{align}\label{eq_Hahncommutant}
\begin{aligned}{}
 K_1&=\frac{1}{2}\left[(N_1+N_2)-(N_3+N_4)\right]\\
 K_2&={L_{12}}^{2}+{L_{13}}^{2}+{L_{14}}^{2}+{L_{23}}^{2}+{L_{24}}^{2}+{L_{34}}^{2}\\
 K_3&=[K_1,K_2],
\end{aligned}
\end{align}
knowing that these operators will satisfy the commutation relations of the Hahn algebra given in \eqref{eq_Hahn2} with
\begin{align}\label{eq_delta12higgs}
\begin{aligned}{}
 \delta_1&=-\frac{\alpha_2}{4}&&=2\left({L_{12}}^{2}-{L_{34}}^{2}\right)H\\
 \delta_2&=\phantom{-}\frac{\alpha_1}{2}&&=\frac{1}{2}H^{2}+4\left({L_{12}}^{2}+{L_{34}}^{2}\right)-2.
\end{aligned}
\end{align}

\section{The embedding of the Hahn algebra into $\mathcal{U}(\mathfrak{su}(1,1))\otimes\mathcal{U}(\mathfrak{su}(1,1))$}\label{sec_embedding}
Let us here indicate how the Hahn algebra is embedded in the tensor product of $\mathcal{U}(\su(1,1))$ with itself. Let $\Delta\,:\,\su(1,1)\to\su(1,1)\otimes\su(1,1)$ be the coproduct homomorphism with $\Delta(J_{\bullet})=J_{\bullet}^{(12)}=J_{\bullet}^{(1)}+J_{\bullet}^{(2)}$ in the superscript notation introduced in Section \ref{sec_oscillators}. Consider the following identification \cite{Granovskii1988,Zhedanov1993}:
\begin{align}\label{eq_Hahn_tensor12}
\begin{aligned}{}
 K_1&=J_0^{(1)}-J_0^{(2)}\\
 K_2&=\Delta(C)=\left[J_0^{(12)}\right]^{2}-J_+^{(12)}J_-^{(12)}-J_0^{(12)},
\end{aligned}
\end{align}
that is $K_2$ is the image of the Casimir element under the coproduct. It is clear that the computation of the overlaps coefficients between the eigenbases of those two operators corresponds to the Clebsch-Gordan problem for $\su(1,1)$.

A simple calculation gives
\begin{align}
 K_2=C^{(1)}+C^{(2)}+2J_0^{(1)}J_0^{(2)}-J_+^{(1)}J_-^{(2)}-J_-^{(1)}J_+^{(2)}.
\end{align}
with $C^{(1)}=C\otimes 1$, $C^{(2)}=1\otimes C$ in keeping with the adopted notation. Let $K_3=[K_1,K_2]$, one finds
\begin{align}
 K_3=-2\left(J_+^{(1)}J_-^{(2)}-J_-^{(1)}J_+^{(2)}\right).
\end{align}
One can now proceed to determine the commutators of $K_3$ with $K_1$ and $K_2$ and one gets:
\begin{align}\label{eq_Hahn_centralext}
\begin{aligned}{}
 [K_3,K_1]&=-2{K_1}^{2}-4K_2+2\left(J_0^{(1)}+J_0^{(2)}\right)^{2}+4\left(C^{(1)}+C^{(2)}\right)\\
 [K_2,K_3]&=-2\{K_1,K_2\}+4\left(J_0^{(1)}+J_0^{(2)}\right)\left(C^{(1)}-C^{(2)}\right).
\end{aligned}
\end{align}
While the first is immediately obtained, a little bit of algebra involving the $\su(1,1)$ commutation relations and its Casimir operator gives the second.

Note that $J_0^{(1)}+J_0^{(2)}$ is central since it commutes with $K_1$ and $K_2$ by construction.

We recognize in \eqref{eq_Hahn_centralext} the commutation relations \eqref{eq_Hahn2} of the (centrally extended) Hahn algebra with
\begin{align}\label{eq_delta12}
\begin{aligned}{}
 \delta_1&=4\left(J_0^{(1)}+J_0^{(2)}\right)\left(C^{(1)}-C^{(2)}\right)\\
 \delta_2&=2\left(J_0^{(1)}+J_0^{(2)}\right)^{2}+4\left(C^{(1)}+C^{(2)}\right).
\end{aligned}
\end{align}
We thus have with the formulas \eqref{eq_Hahn_tensor12}, the embedding of the Hahn algebra in \mbox{$\mathcal{U}(\su(1,1))\otimes\mathcal{U}(\su(1,1))$}. What relation this has to do with the commutant picture will be adressed next.

\section{The Howe duality connection}\label{sec_Howe}
We shall now indicate that the two descriptions of the Hahn algebra presented in Section \ref{sec_Higgs} and \ref{sec_embedding} can be connected through Howe duality. It is known (see in particular \cite{Rowe2012}) that there is a pairing between the representations of $\os(4)$ and $\su(1,1)$ that act in a mutually commuting way (see \eqref{eq:dualact}) on the state space of the four-dimensional harmonic oscillator. We shall exploit this to show that the embedding of the Hahn algebra in the double tensor product of the universal enveloping algebra of one algebra of the pair, $\su(1,1)$, is in duality with the commutant (in the universal algebra of $\us(4)$) of the $\os(2)\oplus\os(2)$ subalgebra of the other algebra of the pair $\os(4)$.

Let us consider the addition of four metaplectic representations \eqref{eq_metaplecticrep} grouped in two pairs, that is take
\begin{align}
 \mathcal{J}_{\bullet}^{(1234)}=\mathcal{J}_{\bullet}^{(12)}+\mathcal{J}_{\bullet}^{(34)}
\end{align}
with
\begin{align}\label{eq_Jpair}
\begin{aligned}{}
 \mathcal{J}_0^{(ij)}&=\frac{1}{2}\left[N_i+N_j+1\right]\\
 \mathcal{J}_+^{(ij)}&=\frac{1}{2}\left(a_i^{\dagger}{}^{2}+a_j^{\dagger}{}^{2}\right)\\
 \mathcal{J}_-^{(ij)}&=\frac{1}{2}\left(a_i{}^{2}+a_j{}^{2}\right).
\end{aligned}
\end{align}
Note that
\begin{align}\label{eq:dualact}
 [L_{ij},\mathcal{J}_\bullet^{(1234)}]=0\qquad\forall i,j=1,\dots,4.
\end{align}
We shall put $\mathcal{J}_{\bullet}^{(12)}$ and $\mathcal{J}_{\bullet}^{(34)}$ in correspondance with the $J_{\bullet}^{(1)}$ and $J_{\bullet}^{(2)}$ of Section \ref{sec_embedding}. In this model,
\begin{align}
 \mathcal{K}_1&=\mathcal{J}_{0}^{(12)}-\mathcal{J}_{0}^{(34)}=\frac{1}{2}\left[(N_1+N_2)-(N_3+N_4)\right]
\end{align}
which is identical with the expression in \eqref{eq_Hahncommutant} for $K_1$ arising from the commutant approach. For $\mathcal{K}_2$ we have
\begin{align}
 \mathcal{K}_2=\mathcal{C}^{(1234)}=\left[\mathcal{J}_0^{(12)}+\mathcal{J}_0^{(34)}\right]^{2}-\left(\mathcal{J}_+^{(12)}+\mathcal{J}_+^{(34)}\right)\left(\mathcal{J}_-^{(12)}+\mathcal{J}_-^{(34)}\right)-\left(\mathcal{J}_0^{(12)}+\mathcal{J}_0^{(34)}\right)
\end{align}
Using \eqref{eq_Jpair}, this becomes
\begin{align}
 \mathcal{K}_2=\frac{1}{4}H^{2}-\frac{1}{2}H-\frac{1}{4}\left(a_1^{\dagger}{}^{2}+a_2^{\dagger}{}^{2}+a_3^{\dagger}{}^{2}+a_4^{\dagger}{}^{2}\right)\left(a_1{}^{2}+a_2{}^{2}+a_3{}^{2}+a_4{}^{2}\right)
\end{align}
and with the help of formulas \eqref{eq_aid2ai2} and \eqref{eq_aid2aj2}, we have
\begin{align}
 \mathcal{K}_2=\frac{1}{4}H^{2}-\frac{1}{2}H-\frac{1}{4}(A_++A_-)-\frac{1}{4}\left((N_1+N_2)^{2}+(N_3+N_4)^{2}-4L_{12}{}^{2}-4L_{34}{}^{2}\right).
\end{align}
This can be rewritten as
\begin{align}\label{eq_K2}
 \mathcal{K}_2=-\frac{1}{4}\left(A_++A_-+\frac{1}{2}D^{2}\right)+\frac{1}{8}H^{2}+L_{12}{}^{2}+L_{34}{}^{2}-\frac{1}{2}
\end{align}
which coincides with the expression that was found when looking for generators commuting with $L_{12}$ and $L_{34}$. Recall that we also found that the expression \eqref{eq_K2} can identically be reexpressed as $\mathcal{K}_2={L_{12}}^{2}+{L_{13}}^{2}+{L_{14}}^{2}+{L_{23}}^{2}+{L_{24}}^{2}+{L_{34}}^{2}$ which makes it also manifest that $\mathcal{K}_2$, calculated as a $\su(1,1)$ Casimir, belongs to the commutant of $\{L_{12},L_{34}\}$ in $\mathcal{U}(\os(4))\subset\mathcal{U}(\us(4))$.

A similar computation shows that the $\su(1,1)$ Casimir for the representation $\mathcal{J}_{\bullet}^{(ij)}$ is given by the square of the corresponding rotation generator in $\os(4)$, namely
\begin{align}
 \mathcal{C}^{(ij)}=L_{ij}{}^{2}-\frac{1}{4}.
\end{align}
It follows that the structure constants become on the basis of \eqref{eq_delta12}:
\begin{align}
\begin{aligned}{}
 \delta_1&=4\left(\mathcal{J}_0^{(12)}+\mathcal{J}_0^{(34)}\right)\left(\mathcal{C}^{(12)}-\mathcal{C}^{(34)}\right)\\
 &=2H\left(L_{12}{}^{2}-L_{34}{}^{2}\right)\\
 \delta_2&=2\left(\mathcal{J}_0^{(12)}+\mathcal{J}_0^{(34)}\right)^{2}+4\left(\mathcal{C}^{(12)}+\mathcal{C}^{(34)}\right)\\
 &=\frac{1}{2}H^{2}+4\left(L_{12}{}^{2}+L_{34}{}^{2}\right)-2
\end{aligned}
\end{align}
in perfect correspondance with \eqref{eq_delta12higgs}. Of course $\mathcal{K}_3=[\mathcal{K}_1,\mathcal{K}_2]$.

Owing to the pairing of the $\su(1,1)$ and $\os(4)$ representations under Howe duality, it is found that the embedding of the Hahn algebra into $\mathcal{U}(\mathfrak{su}(1,1))\otimes\mathcal{U}(\mathfrak{su}(1,1))$ leads to its description as a commutant in $\mathcal{U}(\us(4))$.

\section{Dimensional reduction and the singular oscillator in two dimensions}\label{sec_reduction}
We shall now carry the dimensional reduction of the four-dimensional isotropic harmonic oscillator under the $O(2)\times O(2)$ action to identify in this way the Higgs/Hahn symmetry of the singular oscillator in the plane.

Make the change of variables
\begin{align}
 x_{2j-1}=\rho_{j}\cos\theta_j,\qquad x_{2j}=\rho_j\sin\theta_j,\qquad j=1,2.
\end{align}
Eliminate the $\theta_i$'s by separating the variables with
\begin{align}
 L_{2j-1,2j}=-\frac{i}{2}\frac{\partial}{\partial\theta_j}.
\end{align}
Take the eigenvalues of this operator equal to $-\tfrac{i}{2}k_j$.
After performing the gauge transformation 
$\mathcal{O}\to\widetilde{\mathcal{O}}=(\rho_1\rho_2)^{1/2}\mathcal{O}(\rho_1\rho_2)^{-1/2}$ one sees that the $\su(1,1)$ operators become:
\begin{align}
\begin{aligned}{}
 \widetilde{\mathcal{J}}_0^{(2i-1,2i)}&=\frac{1}{4}\left[-\frac{\partial^{2}}{\partial\rho_i{}^{2}}-\frac{a_i}{\rho_i{}^{2}}+\rho_i{}^{2}\right],\\
 \widetilde{\mathcal{J}}_\pm^{(2i-1,2i)}&=\frac{1}{4}\left[\left(\rho_i\mp\frac{\partial}{\partial\rho_i}\right)^{2}+\frac{a_i}{\rho_i{}^{2}}\right],
\end{aligned}
 \qquad\qquad a_i=k_i{}^{2}+\frac{1}{4},\qquad i=1,2.
\end{align}
The Hamiltonian of the singular oscillator in two dimensions is thus given by
\begin{align}
 \widetilde{H}=2\left[\widetilde{\mathcal{J}}_0^{(12)}+\widetilde{\mathcal{J}}_0^{(34)}\right]=-\frac{1}{2}\left(\frac{\partial^{2}}{\partial\rho_1{}^{2}}+\frac{\partial^{2}}{\partial\rho_2{}^{2}}\right)+\frac{1}{2}\left(\rho_1{}^{2}+\rho_2{}^{2}-\frac{a_1}{\rho_1{}^{2}}-\frac{a_2}{\rho_2{}^{2}}\right).
\end{align}
The constants of motion are clearly
\begin{align}
\begin{aligned}{}
 K_1&=\widetilde{\mathcal{J}}_0^{(12)}-\widetilde{\mathcal{J}}_0^{(34)}\\
 K_2&=\widetilde{C}^{(1234)}\\
 K_3&=[K_1,K_2].
\end{aligned}
\end{align}

We know from our construction that these will close to form the Hahn algebra. The (reduced) Casimir $\widetilde{C}^{(1234)}=\left(\widetilde{J}_0\right)^{2}-\widetilde{J}_+\widetilde{J}_--\widetilde{J}_0$, with $\widetilde{J}_\bullet=\widetilde{\mathcal{J}}_\bullet^{(12)}+\widetilde{\mathcal{J}}_\bullet^{(34)}$ is easily computed and one finds
\begin{align}
\begin{aligned}
 K_2&=-\frac{1}{4}\left[\left(\rho_1\frac{\partial}{\partial\rho_2}-\rho_2\frac{\partial}{\partial\rho_1}\right)^{2}+a_1\left(\frac{\rho_2{}^{2}}{\rho_1{}^{2}}+1\right)+a_2\left(\frac{\rho_1{}^{2}}{\rho_2{}^{2}}+1\right)+1\right],\\
 K_3&=\frac{1}{4}\left[\left(2\rho_1\frac{\partial}{\partial\rho_1}+1\right)\left(\frac{\partial^{2}}{\partial\rho_2{}^{2}}+\rho_2{}^{2}+\frac{a_2}{\rho_2{}^{2}}\right)-\left(2\rho_2\frac{\partial}{\partial\rho_2}+1\right)\left(\frac{\partial^{2}}{\partial\rho_1{}^{2}}+\rho_1{}^{2}+\frac{a_1}{\rho_1{}^{2}}\right)\right].
\end{aligned}
\end{align}
This approach, which combines the commutant viewpoint via the dimensional reduction under the torus group action and the $\su(1,1)$ embedding through the metaplectic representation, provides an alternative and straightforward way of showing that the Hahn algebra is the symmetry algebra of the singular oscillator.

\section{Conclusion}\label{sec_conclusion}
This paper has provided a synthetic description of the Higgs and Hahn algebras in light of Howe duality. With the understanding that the Higgs and the (discrete) Hahn algebras are isomorphic, we have shown that this algebra can be viewed as a commutant in $\mathcal{U}(\us(4))$. It has also been recalled that it can be embedded in the tensor product of $\mathcal{U}(\su(1,1))$ with itself. The two approaches have been linked in view of the fact that $\os(4)$ and $\su(1,1)$ form a dual pair on the state space of the harmonic oscillator in four dimensions. This has also provided context to identify the Hahn symmetry of the singular oscillator in two dimensions through dimensional reduction.

In this respect, one might think of obtaining the higher rank Hahn algebras and by that token the symmetries of the singular oscillator in higher dimensions, by considering the commutant of the sum of $n$ $\os(2)$'s in $\mathcal{U}(\us(2n))$. Take for instance $n=3$. the resulting commutant in  $\mathcal{U}(\us(6))$ would have as subalgebras two Hahn algebras associated to the $(\olin{12})$ and $(\olin{23})$ coordinate sectors as well as the Racah algebra also, since we know \cite{Gaboriaud2018} it is the commutant of $\os(2)\oplus\os(2)\oplus\os(2)$ in  $\mathcal{U}(\os(6))\subset\mathcal{U}(\us(6))$. The entire mixed Hahn-Racah algebra will be an interesting deformation of $\su(3)$. Its analysis would certainly warrant particular attention as this algebra will encompass in particular the properties of the connection coefficients for the various separated solutions of singular oscillators in higher dimensions \cite{Genest2014e,Genest2014d}. We plan to return to this question from this angle.

We would also wish to determine if some Howe duality operates in the case of the algebras, like the Askey-Wilson one, associated to $q$-polynomials. Examining the $q$-Hahn algebra to that end in the wake of the present study might prove illuminating and is in our plans.

\ack
The authors would like to thank E. Ragoucy and P. Sorba for informative discussions.
LV wishes to acknowledge the hospitality of the CNRS and of the LAPTh in Annecy where part of this work was done. 
JG holds an Alexander-Graham-Bell scholarship from the Natural Science and Engineering Research Council (NSERC) of Canada. 
The research of LV is supported in part by a Discovery Grant from NSERC.
SV enjoys a Neubauer No Barriers scholarship at the University of Chicago and benefitted from a Metcalf internship. 
The work of AZ is supported by the National Foundation of China (Grant No. 11771015). 

\section*{References}
\bibliographystyle{unsrtinurl} 
\bibliography{citationsHHA.bib}

\begin{thebibliography}{10}

\bibitem{Higgs1979}
P.~W. Higgs.
\newblock {Dynamical symmetries in a spherical geometry I}.
\newblock {\em J. Phys. A Math. Gen}, 12(3):309, 1979.

\bibitem{Letourneau1995}
P.~L{\'{e}}tourneau and L.~Vinet.
\newblock {Superintegrable Systems: Polynomial Algebras and Quasi-Exactly
  Solvable Hamiltonians}.
\newblock {\em Ann. Phys. (N. Y).}, 243(1):144--168, 1995.

\bibitem{Genest2014}
V.~X. Genest, L.~Vinet, and A.~Zhedanov.
\newblock {The Racah algebra and superintegrable models}.
\newblock {\em J. Phys. Conf. Ser.}, 512(1):012011, 2014.

\bibitem{Granovskii1991}
Y.~I. Granovskii, A.~S. Zhedanov, and I.~M. Lutzenko.
\newblock {Quadratic algebra as a "hidden" symmetry of the Hartmann potential}.
\newblock {\em J. Phys. A. Math. Gen.}, 24(16):3887--3894, 1991.

\bibitem{Granovskii1992}
Y.~Granovskii, I.~Lutzenko, and A.~Zhedanov.
\newblock {Mutual integrability, quadratic algebras, and dynamical symmetry}.
\newblock {\em Ann. Phys. (N. Y).}, 217(1):1--20, 1992.

\bibitem{Zhedanov1992}
A.~Zhedanov.
\newblock {The "Higgs algebra" as a quantum deformation of
  {$\mathfrak{su}(2)$}}.
\newblock {\em Mod. Phys. Lett. A}, 07(06):507--512, 1992.

\bibitem{Howe1987}
R.~Howe.
\newblock {Dual pairs in physics: harmonic oscillators, photons, electrons,
  singletons}.
\newblock In M.~Flato, P.~Sally, and G.~Zuckerman, editors, {\em Proceedings,
  Appl. Gr. Theory Phys. Math. Phys.}, chapter~6, pages 179--206. AMS, Chicago,
  1987.

\bibitem{Howe1989}
R.~Howe.
\newblock {Remarks on Classical Invariant Theory}.
\newblock {\em Trans. Am. Math. Soc.}, 313(2):539--570, 1989.

\bibitem{Rowe2012}
D.~J. Rowe, M.~J. Carvalho, and J.~Repka.
\newblock {Dual pairing of symmetry and dynamical groups in physics}.
\newblock {\em Rev. Mod. Phys.}, 84(2):711--757, 2012.

\bibitem{Granovskii1988}
Y.~A. Granovskii and A.~S. Zhedanov.
\newblock {Nature of the symmetry group of the {$6j$}-symbol}.
\newblock {\em J. Exp. Theor. Phys.}, 94:49--54, 1988.

\bibitem{Zhedanov1993}
A.~S. Zhedanov.
\newblock {Hidden symmetry algebra and overlap coefficients for two ring-shaped
  potentials}.
\newblock {\em J. Phys. A. Math. Gen.}, 26(18):4633--4641, 1993.

\bibitem{Koekoek2010}
R.~Koekoek, P.~A. Lesky, and R.~F. Swarttouw.
\newblock {\em {Hypergeometric Orthogonal Polynomials and Their
  {$q$}-Analogues}}.
\newblock Springer Monographs in Mathematics. Springer Berlin Heidelberg, 2010.

\bibitem{Vinet2018}
L.~Vinet and A.~Zhedanov.
\newblock {The Heun operator of Hahn type}.
\newblock {\em Proc. Am. Math. Soc.}, 2018.
\newblock \href {http://arxiv.org/abs/1808.00153v1}
  {\path{arXiv:1808.00153v1}}.

\bibitem{Bonatsos1995}
D.~Bonatsos, C.~Daskaloyannis, and P.~Kolokotronis.
\newblock {Generalized deformed {$\mathfrak{su}(2)$} algebras, deformed
  parafermionic oscillators and finite W-algebras}.
\newblock {\em Mod. Phys. Lett. A}, 10:2197, 1995.

\bibitem{Leinaas1993}
J.~M. Leinaas and J.~Myrheim.
\newblock {Heisenberg Quantization for Systems of Identical Particles}.
\newblock {\em Int. J. Mod. Phys. A}, 8:3649--3695, 1993.

\bibitem{Bowcock1994}
P.~Bowcock.
\newblock {Representation theory of a W-algebra from generalised DS reduction}.
\newblock 1994.
\newblock \href {http://arxiv.org/abs/hep-th/9403157}
  {\path{arXiv:hep-th/9403157}}.

\bibitem{Barbarin1995}
E.~Barbarin, E.~Ragoucy, and P.~Sorba.
\newblock {Finite W-algebras and intermediate statistics}.
\newblock {\em Nucl. Phys. B}, 442:425--443, 1995.

\bibitem{DeBoer1996}
J.~{De Boer}, S.~Brook, F.~Harmsze, and T.~Tjin.
\newblock {Non-linear Finite W-Symmetries and Applications in Elementary
  Systems}.
\newblock {\em Phys. Reports - Rev. Sect. Phys. Lett.}, 272:139--214, 1996.
\newblock \href {http://arxiv.org/abs/hep-th/9503161v1}
  {\path{arXiv:hep-th/9503161v1}}.

\bibitem{Genest2014a}
V.~X. Genest, L.~Vinet, and A.~Zhedanov.
\newblock {Superintegrability in Two Dimensions and the Racah–Wilson
  Algebra}.
\newblock {\em Lett. Math. Phys.}, 104(8):931--952, 2014.

\bibitem{DeBie2017a}
H.~{De Bie}, V.~X. Genest, W.~van~de Vijver, and L.~Vinet.
\newblock {A higher rank Racah algebra and the {$\mathbb{Z}_2^n$}
  Laplace–Dunkl operator}.
\newblock {\em J. Phys. A Math. Theor.}, 51(2):025203, 2017.

\bibitem{Gaboriaud2018}
J.~Gaboriaud, L.~Vinet, S.~Vinet, and A.~Zhedanov.
\newblock {The Racah algebra as a commutant and Howe duality}.
\newblock {\em J. Phys. A Math. Theor. Lett.}, 2018.
\newblock \href {http://arxiv.org/abs/1808.05261v1}
  {\path{arXiv:1808.05261v1}}.

\bibitem{Gaboriaud2018a}
J.~Gaboriaud, L.~Vinet, S.~Vinet, and A.~Zhedanov.
\newblock {The generalized Racah algebra as a commutant}.
\newblock 2018.
\newblock \href {http://arxiv.org/abs/1808.09518v1}
  {\path{arXiv:1808.09518v1}}.

\bibitem{Tsujimoto2012}
S.~Tsujimoto, L.~Vinet, and A.~Zhedanov.
\newblock {Dunkl shift operators and Bannai–Ito polynomials}.
\newblock {\em Adv. Math. (N. Y).}, 229(4):2123--2158, 2012.

\bibitem{DeBie2015}
H.~{De Bie}, V.~X. Genest, S.~Tsujimoto, L.~Vinet, and A.~Zhedanov.
\newblock {The Bannai-Ito algebra and some applications}.
\newblock {\em J. Phys. Conf. Ser.}, 597:012001, 2015.

\bibitem{DeBie2016a}
H.~{De Bie}, V.~X. Genest, and L.~Vinet.
\newblock {The {$\mathbb{Z}_2^n$} Dirac–Dunkl operator and a higher rank
  Bannai–Ito algebra}.
\newblock {\em Adv. Math. (N. Y).}, 303:390--414, 2016.

\bibitem{Gaboriaud2018b}
J.~Gaboriaud, L.~Vinet, S.~Vinet, and A.~Zhedanov.
\newblock {The dual pair {$Pin(2n)\otimes\mathfrak{osp}(1|2)$}, the Dirac
  equation and the Bannai–Ito algebra}.
\newblock {\em Nucl. Phys. B}, 937:226--239, 2018.

\bibitem{Genest2014e}
V.~X. Genest, L.~Vinet, and A.~Zhedanov.
\newblock {The Dunkl oscillator in three dimensions}.
\newblock {\em J. Phys. Conf. Ser.}, 512(1):012010, 2014.

\bibitem{Genest2014d}
V.~X. Genest and L.~Vinet.
\newblock {The multivariate Hahn polynomials and the singular oscillator}.
\newblock {\em J. Phys. A Math. Theor.}, 47(45):455201, 2014.

\end{thebibliography}

\end{document}